\documentclass[prl,twocolumn,showpacs,superscriptaddress]{revtex4}
\usepackage{graphicx}
\usepackage{amsmath}
\usepackage{bm}
\date{\today}
\begin{document}
\title{Velocity of domain-wall motion induced by electrical current in a ferromagnetic semiconductor (Ga,Mn)As} 
\author{M. Yamanouchi}
\affiliation{Laboratory for Nanoelectronics and Spintronics, Research Institute of Electrical Communication, Tohoku University, Katahira 2-1-1, Aoba-ku, Sendai 980-8577, Japan}
\author{D. Chiba}
\affiliation{ERATO Semiconductor Spintronics Project, Japan Science and Technology Agency, Japan}
\affiliation{Laboratory for Nanoelectronics and Spintronics, Research Institute of Electrical Communication, Tohoku University, Katahira 2-1-1, Aoba-ku, Sendai 980-8577, Japan}
\author{F. Matsukura}
\affiliation{Laboratory for Nanoelectronics and Spintronics, Research Institute of Electrical Communication, Tohoku University, Katahira 2-1-1, Aoba-ku, Sendai 980-8577, Japan}
\affiliation{ERATO Semiconductor Spintronics Project, Japan Science and Technology Agency, Japan}
\author{T. Dietl}
\affiliation{Institute of Physics, Polish Academy of Sciences, PL-02668 Warszawa, Poland; Institute of Theoretical Physics, Warsaw University, Poland}
\affiliation{Laboratory for Nanoelectronics and Spintronics, Research Institute of Electrical Communication, Tohoku University, Katahira 2-1-1, Aoba-ku, Sendai 980-8577, Japan}
\affiliation{ERATO Semiconductor Spintronics Project, Japan Science and Technology Agency, Japan}
\author{H. Ohno}
\affiliation{Laboratory for Nanoelectronics and Spintronics, Research Institute of Electrical Communication, Tohoku University, Katahira 2-1-1, Aoba-ku, Sendai 980-8577, Japan}
\affiliation{ERATO Semiconductor Spintronics Project, Japan Science and Technology Agency, Japan}
\email{ohno@riec.tohoku.ac.jp}

\begin{abstract}
Current-induced domain-wall motion with velocity spanning over five orders of magnitude up to 22~m/s has been observed by magneto-optical Kerr effect in (Ga,Mn)As with perpendicular magnetic anisotropy. The data are employed to verify theories of spin-transfer by the Slonczewski-like mechanism as well as by the torque resulting from spin-flip transitions in the domain-wall region. Evidence for domain-wall creep at low currents is found.
\end{abstract}
\pacs{72.25. -b, 75.60.Jk, 75.50.Pp, 85.75.-d}
\maketitle

The manipulation of magnetization by electrical means without external magnetic fields involves outstanding physical phenomena not fully understood by current theories and at the same time it is technologically important because of possible power reduction for magnetization reversal in high-density magnetic memories. One of the well-known schemes for electrical manipulation of magnetization reversal is the injection of spin-polarized current into magnetic multilayer nanopillars \cite{Slon96}, which has been demonstrated in a number of metal systems \cite{Fert04} as well as for a tunnel junction of a ferromagnetic p-type semiconductor (Ga,Mn)As \cite{Chib04,Mats02}. Another scheme which is of focus here is the magnetic domain-wall (DW) displacement by the injection of electrical current, the theory of which has been developed since 1980's \cite{Berg79}, and is now investigated extensively from both experimental \cite{Yama04,Frei85,Yamag04,Sait04} and theoretical \cite{Barn04,
Tata04,Zhan04,Thia05,Tata05} points of view. Recent experiments on ferromagnetic metal NiFe nanowires at room temperature showed that DW can be moved by the application of pulsed current with the density $j$ of $10^7 \sim 10^8$~A/cm$^2$ \cite{Yamag04} or by ac current with $j \sim 10^6$~A/cm$^2$ and frequency in the MHz range \cite{Sait04}. For (Ga,Mn)As, we demonstrated that DW movement by current pulses is possible with $j \sim 10^5$~A/cm$^2$ around 80~K \cite{Yama04}.

In this Letter, we present studies on dependence of DW motion on the current density and temperature in (Ga,Mn)As. We have observed DW velocities spanning over five decades, which makes it possible to examine various mechanisms accounting for the current-induced DW displacement. In particular, we show that DW motion we have observed is not caused by the Oersted field of the current circulating around the DW, a drag mechanism considered in the pioneering work of Berger \cite{Berg79}. Instead, we demonstrate that the spin-transfer regime \cite{Berg84,Tata04} has been reached at high current densities in (Ga,Mn)As. We show that the recent theory of this mechanism \cite{Tata04}, developed within the s-d--type model and thus directly applicable to the hole-mediated ferromagnetic (Ga,Mn)As, describes experimental magnitudes of both the critical currents $j_{\mbox{\tiny{C}}}$ and the DW velocities $v_{\mbox{\tiny{eff}}}$ within a factor of two. We examine also the scaling properties
 of $v_{\mbox{\tiny{eff}}}$ below $j_{\mbox{\tiny{C}}}$ \cite{Tata05}, and suggest that spin-current assisted DW creep is involved \cite{Leme98}. Furthermore, we discuss our results in view of recent theories \cite{Zhan04,Thia05} that link the low-current effects to the presence of an additional torque brought about by non-adiabatic carrier transfer across DW. 

The sample grown by molecular beam epitaxy consists of Ga$_{0.955}$Mn$_{0.045}$As (30~nm)/ In$_{0.22}$Al$_{0.78}$As (75~nm)/ In$_{0.18}$Al$_{0.82}$As (75~nm)/ In$_{0.13}$Al$_{0.87}$As (75~nm)/ In$_{0.065}$Al$_{0.935}$As (75~nm)/ semi-insulating GaAs (001) substrate. A stack of (In,Al)As buffer layers is employed to make magnetic easy axis of (Ga,Mn)As parallel to the growth direction by the controlled lattice strain \cite{Shen97}, and its stepped graded composition reduces surface roughness due to crosshatch dislocations resulting from a lattice mismatch to the substrate \cite{Mish95}, which may disturb uniform DW motion. After the formation of a 5~$\mu$m wide current channel along [-110] direction, a  10~nm surface layer is etched away from a part of the channel. As shown in Fig.~1(a) we use 60~$\mu$m long etched region (I) and 20~$\mu$m non-etched region (II) for domain structure observation. The rest of the channel is covered by Au/Cr electrodes, which reduces the device
resistance $R$ between the two Au/Cr electrodes to about 22~k$\Omega$ at $\sim 100$~K. 

\begin{figure}[]
\includegraphics[width=3.3in]{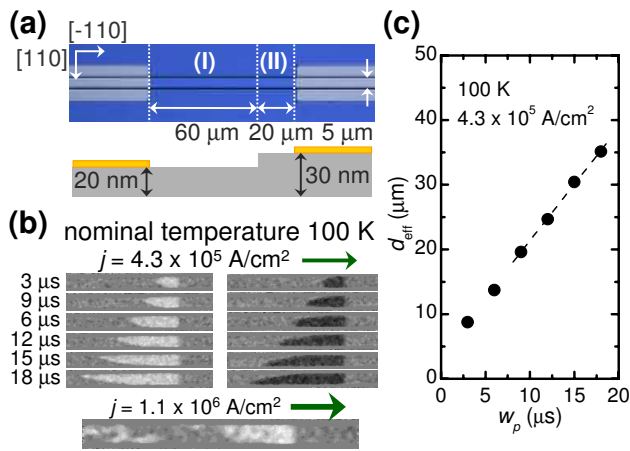}
\caption{[Color online] (a) Layout of the device showing 5~$\mu$m mesa and step for domain wall (DW) pinning in perpendicular magnetic anisotropy (Ga,Mn)As film. (b) 7 $\mu$m wide magneto-optical images with 5 $\mu$m mesa in the center show that DW moves in the opposite direction to current independent of the initial magnetization orientation, and that DW displacement is proportional to  pulse duration (c). The lowest panel in (b) shows destruction of ferromagnetic phase by Joule heating.} \label{fig:1}
\end{figure}

The ferromagnetic transition temperature $T_{\mbox{\tiny{C}}}$ of regions (I) and (II) is determined by a magneto-optical Kerr effect (MOKE) microscope to be 112~K and 115~K, respectively, above which no domain structures are detectable. The difference in $T_{\mbox{\tiny{C}}}$ gives different coercive force in each region, which allows us to initialize the DW position at the boundary of the two regions by an external magnetic field. Once DW is created in the channel, DW can be moved back to the step boundary by the current with a good reproducibility \cite{Yama04}. For the uniaxial magnetic anisotropy energy $K_u$ and magnetic stiffness $A_s$ corresponding to tensile strain and Mn concentration in question we evaluate the width of the Bloch wall to be $\delta_W = \pi(A_s/K_u)^{1/2} \approx 17$~nm \cite{Diet01b}, which for actual values of in-plane anisotropy energies should be energetically more stable than the N\'eel wall. 

We find that for the present arrangement the transition and trailing times of the pulses are about 500~ns, and thus we choose the minimum current pulse width to be 1~$\mu$s; the maximum is set to 800~ms. During the application of the pulse, we screen the device from ambient light to avoid the effect of photoconductivity in the buffer layer. For the observation of the domain structure we use MOKE microscope with 546~nm light. In order to enhance the image contrast, we register differential images before and after the application of the pulse, {\it i.e.}, the brightness of the image changes only in the area where the reversal of magnetization occurred by the current injection. In this way we obtain the images shown in Fig.~1(b), where the increase of white (black) area corresponds to the increase of the area with positive (negative) magnetization direction with respect to the initial magnetization configuration (DW at the boundary). We measure the reversed area ({\it i.e.}, the area
 swept by DW) $A_d$ with pulses of various amplitudes $j$ and widths $w_p$ at nominal temperatures of $T_a$ = 92, 94, 96, 100, and 104~K. The effective displacement of DW $d_{\mbox{\tiny{eff}}}$ is determined as a ratio of $A_d$ to the channel width $w$. In order to avoid electric breakdown, the maximum $j$ is restricted to $1.3\times10^6$~A/cm$^2$. Figure 1(b) presents the dependence of MOKE images on the pulse duration $w_p$ at $j = 4.3\times10^5$~A/cm$^2$ and $T_a = 100$~K. The left panel corresponds to the initial configuration with magnetization pointing down (negative $M$) in region (I) and positive $M$ in region (II). The right panel is for the opposite initial configuration, which results in the reversed brightness of the DW area swept by the current injection. We have found that DW always moves in the opposite sense to the current direction independently of the initial magnetization orientation. 

There are two sources of the Oersted field brought about by current, which can lead to DW motion. First, in a thin uniform conductor, $t \ll w$, the field generated by the current is concentrated on the two edges, and its averaged component over the thickness $t$ is $H_z = \pm jt[3+2\ln(w/t)]/4\pi$, reaching the magnitude $\mu_o|H_z| \approx 0.4$~mT in the present experiment. However, if this were the source of DW motion, the direction of motion would have depended on the initial magnetization configuration, in contrast to our observations. Second, the current and magnetization produce a transverse (anomalous) Hall electric field that changes its sign on crossing DW. This generates an additional current that circulates around DW \cite{Part74}, and induces a magnetic field $H_z'$ reaching a maximum value in the DW center. Averaging over the DW width $\delta_W$ we obtain $H_z' = bjt\tan\theta_H$, where the sign corresponds to a positive direction of magnetization in the source
 contact and $b \approx 2.0$, independent of the Hall angle $\theta_H$ and $w$ under our experimental conditions, $|\tan\theta_H| \le 0.1$ and $\delta_W \ll w$. Because $\theta_H > 0$ in the studied layers, this hydromagnetic DW drag force moves DW in the direction of the current, again in contradiction to our findings.

\begin{figure*}[]
\includegraphics[width=7in]{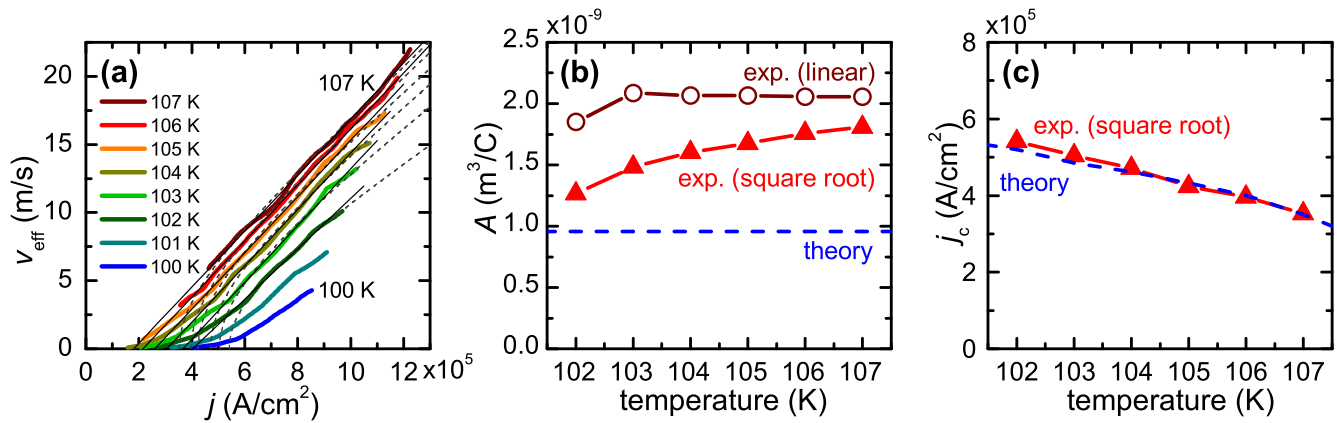}
\caption{[Color online] (a) DW velocity as a function of current at various device temperatures. Thin line and broken thin line show fitted linear and square root dependencies of velocity on current, respectively.  Efficiency factor $A$ (b) and critical current density $j_{\mbox{\tiny{C}}}$ (c) resulting from these two fits (empty and full symbols, respectively). Broken lines are theoretically calculated assuming that spin current polarization is equal to thermodynamic spin polarization.} \label{fig:2}
\end{figure*}

In view of the above considerations, we turn to the description of our results in terms of spin-transfer. Since the sign of p-d exchange integral $\beta$ between the hole carriers and localized Mn spins is negative, a simple application of spin momentum conservation implies that the DW is expected to move in the opposite direction to the electric current, as observed. Conversely, our findings can be taken as an experimental evidence for the antiferromagnetic sign of the p-d coupling in (Ga,Mn)As.

Figure 1(c) shows the dependence of DW displacement $d_{\mbox{\tiny{eff}}}$ on pulse width $w_p$ obtained from Fig.~1(b). For longer $w_p$, $d_{\mbox{\tiny{eff}}}$ increases linearly with $w_p$; to reduce the possible experimental errors accompanied by the region near the stepped boundary, we define the effective velocity $v_{\mbox{\tiny{eff}}}$ as the slope d$d_{\mbox{\tiny{eff}}}$/d$w_p$ for $d_{\mbox{\tiny{eff}}} > 15$~$\mu$m. We also note that the swept area has a wedge shape, and its edge side is reversed for the reversed initial $M$ configurations. Two effects can work together to produce such a behavior. First, an asymmetric DW motion can be induced by the Oersted field $H_z$ that is oriented in the opposite sense at the two channel edges. Second, the current at the edges in the DW region is either enhanced or reduced by the jump in the Hall electric field, depending on the magnetization configuration.

In order to take into account the device temperature increase $\Delta T$ due to Joule heating, we compared the temperature dependence of the device resistance $R$ measured at low $j$ of $5\times 10^3$~A/cm$^2$ and $R$ during the application of the pulse as a function of $j$. We find that $\Delta T$ is more sensitive to $j$ than to $w_p$ and that the values of $\Delta T$ determined at different $T_a$ by using $R(T)$ as a thermometer collapse into a single curve, $\Delta T [{\mbox K}]  = 7.63\times 10^{-12}j^2[\mbox{A/cm}^2] + 2.86\times 10^{-6}j[\mbox{A/cm}^2]$. The validity of this $\Delta T$ determination is supported by a multi-domain structure observed after the application of pulses with high $j$, where $T_a +\Delta T$ is above $T_{\mbox{\tiny{C}}}$, as shown in Fig.~1(b). This $\Delta T$ limits the temperature range at which DW motion can be observed. We hereafter use the calibrated temperature $T = T_a +\Delta T$ to describe our results. A maximum value of $v_{\mbox{\tiny{eff}}}$
 found in the present measurements is 22~m/s at $T = 107$~K and $j = 1.2\times 10^6$~A/cm$^2$, which is about 4 times greater than that reported for a NiFe nanowire \cite{Yamag04}.

By employing the procedure outlined above, we obtain $v_{\mbox{\tiny{eff}}}$ vs.~$j$  at various $T$ collected in Fig.~2(a), where $v_{\mbox{\tiny{eff}}}$ is seen to increase almost linearly with $j$. We determine the spin-transfer efficiency factor $A$ and the critical current density for DW motion $j_{\mbox{\tiny{C}}}$ assuming either a linear dependence with the slope $A$ and  a threshold current density related to defect pinning as proposed by Barnes and Maekawa \cite{Barn04}, or that resulting from theory of Tatara and Kohno \cite{Tata04}, $v_{\mbox{\tiny{eff}}}= A(j^2 - j_{\mbox{\tiny{C}}}^2)^{1/2}$. Their models constitute a continuous version of Slonczewski's approach \cite{Slon96}, and are developed in the adiabatic limit which means that DW does not introduce any additional carrier scattering and that carrier spin polarization tracks local magnetization $\bm M$ of the Mn spins. The latter is satisfied under our experimental conditions, as the hole precession time in the molecular field of
 the Mn spins, $\tau_{ex} = \hbar g\mu_B/|\beta| M$ is shorter than both dwell time of the holes diffusing across DW, $\tau_D = \delta_W^2/D$, where $D$ is the diffusion constant, and $\tau_{sf}$ is spin-flip scattering time limited by spin-orbit interactions and spin-mixing by non-colinearity of magnetization inside DW. Within this scenario, neglecting damping and pinning, $A = g\mu_BP/2eM$  \cite{Tata04, Barn04} and $j_{\mbox{\tiny{C}}} = 2eK\delta_W/\pi \hbar P$ \cite{Tata04}, where $P$ is the spin-current polarization. In this regime, according to the Landau-Lifshitz-Gilbert (LLG) equation, we deal with DW motion accompanied by in-plane Mn spin precession with an average frequency $\omega_{\varphi} = \pi \alpha_Gv_{\mbox{\tiny{eff}}}/\delta_W$, reaching ~81~MHz if the Gilbert damping constant $\alpha_G = 0.02$~\cite{Sino04} and $v_{\mbox{\tiny{eff}}} = 22$~m/s. 

To interpret the values of $A$ and $j_{\mbox{\tiny{C}}}$ we assume $P$ to be equal to the thermodynamic spin polarization $P_s$ which, according to the p-d Zener model \cite{Diet01a}, is given by $P_s = 6k_BT_{\mbox{\tiny{C}}}M/[(S + 1)p\beta M_S]$ in the mean-field approximation. To evaluate $P_s$ the hole concentration is taken as $p = 4\times 10^{20}$~cm$^{-3}$; $\beta = -0.054$~eV nm$^3$ \cite{Diet01a} and saturation magnetization $M_S$ is calculated, while spontaneous magnetization $M = M(T)$ has been measured on a larger area sample cleaved from the same wafer. The energy density $K$ of magnetic anisotropy associated with the rotation of the DW spins in the plane perpendicular to the layer easy axis is evaluated as the stray field energy density $\mu_oM^2/(2+4\delta_w/\pi t)$ of the spins in the DW plane, whose magnitude is five times greater than the in-plane crystal magnetic anisotropy constants evaluated from ferromagnetic-resonance spectra of similar films \cite{Yama05}. In view
of the uncertainty in $p$ and $\beta$ as well as in the relation between $P$ and $P_s$ in the case of the complex valence band structure, we conclude from Figs.~2(b,c) that our experimental results corroborate the spin-transfer theory \cite{Tata04}. Furthermore, as can be seen in Fig.~2(a) (see also Fig.~3), we observe non-zero DW velocities below $j_{\mbox{\tiny {C}}}$, where under stationary conditions the torque exerted by the flow of spin-polarized carriers is compensated by the deflection of the DW spins from the equilibrium orientation. This subthreshold effect is particularly noticeable at high temperatures, and may contribute to the enhanced $A$ for $j > j_{\mbox{\tiny {C}}}$. The existence of such a contribution may explain a different slope of experimental and theoretical values $A(T)$ visible in Fig.~2(b).

We assign the smearing of the critical behavior near $j_{\mbox{\tiny{C}}}$ and the associated enhanced velocities to non-zero temperatures of our experiments. As shown in Fig.~3, we find that in the range $j \lesssim j_{\mbox{\tiny {C}}}$, $v_{\mbox{\tiny{eff}}}(j,T)$ obeys over many decades an empirical scaling law, $\ln v_{\mbox{\tiny{eff}}} = a(T) -b(T)j^{-\nu}$, where $\nu = 0.5\pm 0.1$; $a(T) \sim (T_{\mbox{\tiny{C}}}-T)$, and $b(T) \sim (T_{\mbox{\tiny{C}}}-T)^{\gamma}$ with $\gamma = 2 \pm 0.5$. Since $\nu < 1$ we rather deal with DW creep \cite{Leme98} than with over-barrier thermal activation that moves DW as a whole \cite{Tata05}. Furthermore, the large magnitude of $\gamma$ indicates that the thermally-assisted effects become particularly strong on approaching $T_{\mbox{\tiny{C}}}$. We thus suppose that in this regime spin-current-induced DW creep is triggered by critical magnetization fluctuations that diminish locally $j_{\mbox{\tiny {C}}}$. 

\begin{figure}[]
\includegraphics[width=3.3in]{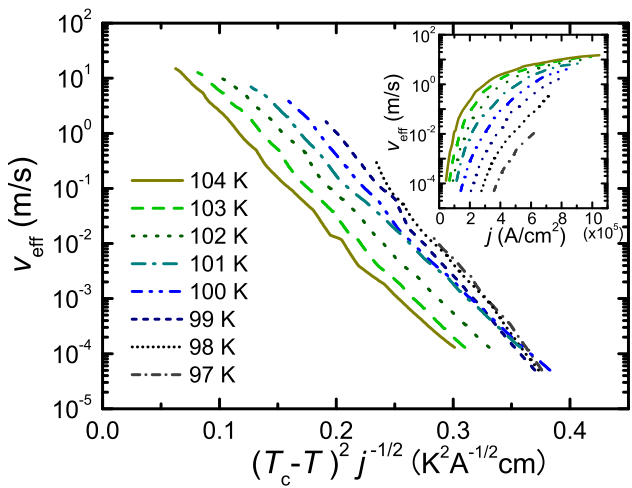}
\caption{[Color online] A test of scaling law for DW creep in (Ga,Mn)As at various temperatures. Inset shows dependence of DW velocity on current density.} \label{fig:3}
\end{figure}

It has recently been suggested \cite{Zhan04,Thia05} that the current-induced DW displacement in metallic wires occurs at $j$ $<$ $j_{\mbox{\tiny {C}}}$ (thus $j$ $>$ $j_{\mbox{\tiny {C}}}$ has not been reached in metallic structures), which has been taken as an evidence for the existence of another current-induced torque for which $v_{\mbox{\tiny{eff}}} = Cj$ \cite{Zhan04,Thia05}. Such a torque may appear in LLG equation if a finite spin-relaxation time $\tau_{sf}$ is taken into account in the carrier spin-diffusion equation inside  DW \cite{Zhan04}. Within this model $C = A\tau_{ex}/\tau_{sf}\alpha_G$. From the subthreshold slope of Fig.~2(a), and with the hydromagnetic force $H_z'$ taken into account, we obtain the upper limit for $C\approx 5\times 10^{-10}$ m$^3$/C which leads to $\tau_{ex}/\tau_{sf}\approx 1\times 10^{-2}$, a value reasonable for (Ga,Mn)As.  This approach is also consistent with the thermally-assisted character of DW motion in the low-current regime under the presence of extrinsic pinning. The above value of $\tau_{ex}/\tau_{sf}$ implies that the relaxation damping does not affect the {\em over}-threshold DW velocity, reduced by a factor $ \{[1+(\tau_{ex}/\tau_{sf})^2](1+\alpha_G^2)\}^{-1}$ \cite{Zhan04} close to 1.

In summary, our work substantiates quantitatively the notion that the current-induced domain-wall (DW) displacement by spin-transfer mechanism results from the Slonczewski-like torque brought about by the decay of the carrier spin-polarization component perpendicular to the local magnetization of the Mn spins.  This mechanism starts to operate when the corresponding torque overcompensates the counter reaction torque generated by the current-induced change in the magnetic anisotropy energy.  Interestingly, we find that the DW displacement is still possible at lower currents owing to spin-current assisted creep. An effect brought about by a torque resulting from spin-flip transitions within the DW region may also contribute in this regime. 
    
The authors thank G. Tatara, A.H. MacDonald, Y. Ohno, K. Ohtani, and T. Kita for useful discussions. This work was supported in part by the IT Program of RR2002 from MEXT, Grant-in-Aids from MEXT/JSPS, Research Fellowship from JSPS, and the 21st Century COE program at Tohoku University.

\end{document}